\documentclass[a4paper]{article}

\usepackage[english]{babel}
\usepackage[utf8]{inputenc}
\usepackage{amsmath}
\usepackage{graphicx}
\usepackage{epsfig}
\usepackage[colorinlistoftodos]{todonotes}
\usepackage[hidelinks]{hyperref}
\usepackage[margin=1.3in]{geometry}
\usepackage{natbib}
\usepackage{breqn}
\usepackage{amssymb}
\usepackage{bm}
\usepackage{bbm}
\usepackage{authblk}

\title{A Survey of Causal Inference Frameworks}

\author[1]{Jingying Zeng}
\author[2]{Run Wang}
\affil[1]{Georgia Institute of Technology}
\affil[2]{Iowa State University}

\date{September 2022}

\begin{document}
\maketitle

\begin{abstract}
Causal inference is a science with multi-disciplinary evolution and applications.
On the one hand, it measures effects of treatments in observational data based on experimental designs and rigorous statistical inference to draw causal statements. One of the most influential framework in quantifying causal effects is the potential outcomes framework. On the other hand, causal graphical models utilizes directed edges to represent causalities and encodes conditional independence relationships among variables in the graphs. A series of research has been done both in reading-off conditional independencies from graphs and in re-constructing causal structures. In recent years, the most state-of-art research in causal inference starts unifying the different causal inference frameworks together. This survey aims to provide a review of the past work on causal inference, focusing mainly on potential outcomes framework and causal graphical models. We hope that this survey will help accelerate the understanding of causal inference in different domains.
\end{abstract}


\section{Introduction}
Causal reasoning is widely used in everyday language interchangeably with correlation or association. Based on observations, one might draw informal conclusions like "Aspirin helps reduce my pain" or "She got a good grade because she is smart". In many cases, "correlation does not imply causation"; however, an action (or manipulation) can cause an effect. Therefore, lots of causal statements are implausible if no explicit intervention being considered. Traditional statistical analysis such as estimation and hypothesis testing typically emphasizes on the associations among variables and bypass the language of causality. Nevertheless, the questions that scientific researchers trying to answer are more related to causality in nature rather than correlation. For instance, will smoking cause lung cancer? Will a decrease in demand of houses cause housing price drop? Can a certain drug prolong survival for cancer patients? Causal inference goes beyond association and further identifies the causal effects when a cause of the effect variables is intervened, which makes it an critical research area across many fields, such as computer science, economics, epidemiology, and social science.

In scientific research, controlled randomized experiment using random assignment mechanisms to assign subjects to different treatment groups, has a long history in serving as the gold standard in establishing causality. In many situations, however, randomized experiments are not feasible nor ethical in practice, so that researchers need to rely on observational data to inference causal relationships. To this end, several research communities have developed various frameworks for causal identification based on observational data.

In history, there are three major origins of causal inference developed separately for different purposes, namely potential outcomes (counterfactuals) community, graphs community, and structural equations community. The potential outcomes framework provides a way to identify causal effects using statistical inference. This framework was first introduced by \citet{neyman1923application} for randomized experiments and generalized by \citet{rubin1974estimating} for observational studies. Later, \citet{holland1986statistics} published an influential paper and named the general potential outcomes method as Rubin Causal Model. The potential outcomes approach associates causality with manipulation applied to units, and compares causal effects of different treatments via their corresponding potential outcomes. It is convenient to use potential outcomes to make statistical inference on single cause-effect pair, and yet it has some drawbacks when the system becomes complicated. On the other hand, graph analysis and structural equations in causal inference traced back to the works in path analysis introduced by geneticist \citet{wright1918nature}. Wright's path analysis combines graphs with linear structural equation models (SEMs) to present causal relationships by directed edges, which is a useful tool to differentiate correlation from causation when the graph structure is given. \citet{pearl1988probabilistic} later relaxed linearity assumption and formalized causal graphical models for presenting conditional independence relations among random variables using directed acyclic graphs (DAGs). A variety of criteria have been developed for reading-off independencies from a given graph, including the most famous theorem, the completeness of d-separation. Nowadays, researchers have been making efforts in unifying the theories of causality at the intersection of those communities \citep{richardson2013single}, and using the causal inference to make estimations more reliable \citep{rojas2018invariant, arjovsky2019invariant}.

\section{Potential Outcomes Framework}
One of the most popular statistical frameworks inferring causal effects is Neyman-Rubin model, also referred to as potential outcomes model. For each unit $i$ from a target population under a binary treatment $Z_i \in \{0,1\}$, the framework quantifies the individual causal effects via comparing the difference between potential outcomes for this unit in both alternative futures, denoted as $\Delta_i = Y_i(1)-Y_i(0)$, where $Y_i(1)$ and $Y_i(0)$ are the responses of the subject $i$ corresponding to treatment or control respectively \citep{neyman1923application, rubin1974estimating, rubin1977assignment, rubin1978bayesian}. Potential outcomes are also referred to as counterfactuals in literature. Even though $\Delta_i$ can never be directly estimated because $Y_i(1)$ and $Y_i(0)$ can never be simultaneously observed for each unit, statisticians shed light on this fundamental problem in causal inference by shifting the target estimand to the average treatment effect (ATE) \citep{holland1986statistics}. Since at most of one of the potential outcomes can be observed and realized, causal inference under potential outcomes framework is intrinsically a missing data problem \citep{rubin1976inference}. In statistical literature, there are generally two versions in defining ATE, namely population ATE (PATE) $\tau_P$ and finite-sample ATE (FATE) $\tau_{FS}$, formulated as follows \citep{imbens2015causal}:
\begin{align}\label{estimand}
\tau_{P} \equiv \mathbb{E}\big[ Y_i(1)-Y_i(0) \big]\\
\tau_{FS} \equiv \sum_{i=1}^{N}\big[ Y_i(1)-Y_i(0) \big]
\end{align}

Both estimands can be viewed as a function of $\{\bm{Y}(0), \bm{Y}(1), \bm{X}, \bm{Z} \}$, where $\bm{X}$ is the observed covariate matrix. However, the distinction between the two is whether
making the assumption that the potential outcomes are stochastic. For PATE, N subjects, each associated with a quadruple $\{ Y_i(0),Y_i(1),Z_i,\bm{X}_i \}$, are viewed as a random sample drawn from a super-population, which induces the fact that the potential outcomes are random variables with a distribution. When FATE is the target estimand, the potential outcomes are treated as a fixed vector and inferences are conditional on this vector. The randomness comes from the treatment assignment $Z_i$'s. With random sampling generated from the target population, SATE is in expectation equal to PATE. SATE is typically of interest in randomized experiments. Conversely, observational studies often utilize PATE as target causal estimand. Intuitively speaking, for instance, an investigator randomly sampled a hundred of subjects from the U.S to test the efficacy of a vaccine. The question that causal inference is trying to answer under finite population is whether this vaccine is effective for this sample of subjects, yet under super population settings, the question shifts to whether the efficacy of this vaccine can be generalized for the whole U.S population.

According to the potential outcomes model of binary treatment, the values of $\big(Y_i^{obs}, Z_i, \bm{X}_i\big)\in \mathbb{R}  \times \{0,1\} \times \mathcal{X} $ of $N$ independent and identically distributed samples were observed, where $Z_i$ and $\bm{X}_i$ denote the treatment assignment and a vector of covariates for each sample unit. By assuming Stable Unit Treatment Value Assumption (SUTVA) (no different forms of treatment level and no interference among sample subjects), the observed outcome is deterministic by the treatment assignment as $Y_i^{obs}=Y_i(Z_i)$, or equivalently, $Y_i^{obs}=Z_iY_i(1) + (1-Z_i)Y_i(0)$ \citep{rubin1980randomization}. Similary, the unobserved potential outcome can be denoted as $Y_i^{miss}=Y_i(1-Z_i)$.

In addition to the observed outcomes, \citet{rubin1990comment} claims that the assignment mechanism, defined as the process of determining which subject receives which treatment level, is also an essential piece of information in inferring reliable causal estimands. In potential outcomes framework, causal estimands are disentangle from the probabilistic models of assignment mechanism, which is the primary difference that distinguishes potential outcomes model from other frameworks \citep{imbens2015causal, frangakis2002principal}. Classical randomized controlled trial is a randomized experiment with a controlled and random assignment mechanism, which guarantees the unconfoundedness such that $ \{ Y_i(0), Y_i(1)\} \perp Z_i | \bm{X}_i$ \citep{rubin1978bayesian}. 

In a randomized experiment, randomization balances both observed and unobserved baseline characteristics \citep{rubin2008objective} so that a simple difference-in-means estimator, also known as Neyman's inference, formulated as 
\begin{equation*}
    \hat{\tau}_{DIM}=\frac{1}{N_1}\sum_{Z_i=1}Y_i - \frac{1}{N_0}\sum_{Z_i=0}Y_i,
\end{equation*}
where $N_1 = \sum_{i=1}^N Z_i$ and $N_0 = \sum_{i=1}^N (1-Z_i)$, can consistently and unbiasedly estimating the ATE and establish credible causal link at the population level \citep{imbens2015causal}. Studying on randomized experiments has attracted much attention from different researchers. \citet{fisher1992statistical} considered randomization test in testing the sharp null hypothesis of whether $Y_i(0)-Y_i(1)=\beta$ $\forall i$, which in practice is often approximated by Monte-Carlo simulation. \citet{aronow2014sharp} constructed a sharper bound on the variance of the difference-in-means estimator. In modern causal inference, theories indicate that appropriately incorporating pre-treatment covariates can increase precision in randomized experiments analyses \citep{li2020rerandomization, ye2020principles}, and many researchers further investigated the use of ordinary least squares regression-adjusted estimates that do not depend on the assumptions of linear model to improve asymptotic accuracy, including the discussions by \citet{freedman2008regression}, \citet{tsiatis2008covariate}, \citet{lin2013agnostic}, and \citet{ye2020principles}. Making inference about causal effects from randomized experiments have been viewed as gold standard, however, randomized experiments can be infeasible and unethical in practice. Therefore, studying on observational data for causality serves as an alternative when it is impossible to conduct randomized experiments.

The biggest challenge for observational studies is that the treatment assignment mechanisms are unknown and the subjects assigned to different groups might systematically differ in some unobserved characteristics. Thus, additional assumptions are necessary to identify causality. To obtain credible causal effects from observational studies, one needs to see how well an observational study emulate a randomization-like scenario. The majority of causal inferences for observational studies are based on the strong ignorability assumption on assignment mechanisms, which requires unconfoundedness and positivity to make it free from latent bias \citep{rosenbaum1983central}. The unconfoundedness assumption assumes that with a set of controlled covariates $\bm{X}_i$ given, the assignment mechanism is independent of the potential outcomes. When one is willing to assume unconfoundedness, an observational study can be considered as a randomized controlled trial and the treatment assignment is random defined by a level of $\bm{x}$. In addition to unconfoundedness, the positivity assumption further requires propensity score, defined as $e(\bm{x}) = \mathbb{P}(Z_i=1|\bm{X}_i = \bm{x})$, to be strictly within 0 and 1 for all $\bm{x}$ in the support of $\bm{X}_i$, so that each unit has positive probability of being assigned to either treatment or controlled group. Without positivity assumption, the probability of a certain subpopulation to be assigned to one of the groups might be zero, and the inference on treatment effects for this subpopulation will depend on extrapolation \citep{imbens2015causal}.

By assuming no unmeasured confounders, one popular way to derive an estimator for the ATE is via regression models. Defining $\mu_z(\bm{x})= \mathbb{E}\big[ Y_i(z)|\bm{X}_i=\bm{x} \big]$ as the outcome regressions among treated and control respectively and $\hat{\mu}_z(\bm{X}_i)$ as the corresponding fitted models, an alternative representation of the PATE is 
\begin{equation*}
    \tau_{P} = \mathbb{E}_{\bm{X}}\big[ \mathbb{E}\big( Y_i(1)|\bm{X} \big) \big] - \mathbb{E}_{\bm{X}}\big[ \mathbb{E}\big( Y_i(0)|\bm{X} \big) \big]
    = \mathbb{E}_{\bm{X}} \{   \mathbb{E}(Y_i^{obs}|Z=1,\bm{X})  - \mathbb{E}_{\bm{X}} (Y_i^{obs}|Z=0,\bm{X})  \}
\end{equation*}
suggesting a regression-based estimator can be constructed as follows  \citep{rubin1979using}:

\begin{equation*}
    \hat{\tau}_{reg} = \hat{\tau}_{reg,1}-\hat{\tau}_{reg,0} =\frac{1}{N}\sum_{i=1}^{N}\big[  \hat{\mu}_1(\bm{X}_i) - \hat{\mu}_0(\bm{X}_i)  \big].
\end{equation*}
The advantage of using regression-based estimator $\hat{\tau}_{reg}$ is that it can easily be estimated by machine learning methods. However, the unbiasedness and consistency of this estimator $\hat{\tau}_{reg}$ are ensured by correct specification on the postulated outcome regression models, which might be hard to guarantee when $\bm{X}$ is high-dimensional due to the curse of dimensionality \citep{glynn2010introduction}.

A common way to generalize randomized controlled trials in observational studies is through propensity scores. One of the most appealing properties of propensity score, showed by \citet{rosenbaum1983central}, is that, if the treatment assignment is unconfounded given $\bm{X}_i$, it is also unconfounded given the propensity score, suggesting that adjusting according to propensity score can remove confounding in observational studies. Once propensity score is estimated, methods such as matching, stratification, and weighting can be further applied to make causal inference.


\subsection{Weighting Methods}	

Most early studies focused on the causal effects of binary treatments. Inverse Propensity Weighting (IPW) is a popular weighting estimator estimating the PATE under strong ignorability assumption. The IPW estimator formulated as follows \citep{rosenbaum1987model, horvitz1952generalization}:
\begin{equation*}
\hat{\tau}_{IPW} = \hat{\tau}_{IPW,1} - \hat{\tau}_{IPW,0} = \frac{1}{N} \sum_{i=1}^{N}\frac{Z_iY_i^{obs}}{\hat{e}(\bm{X}_i)} -  \frac{1}{N} \sum_{i=1}^{N}\frac{(1-Z_i)Y_i^{obs}}{1-\hat{e}(\bm{X}_i)}.
\end{equation*} 
The IPW estimator $\hat{\tau}_{IPW}$ is the same as the regression estimator $\hat{\tau}_{reg}$ when $\bm{X}$ is discrete, but when $\bm{X}$ is continuous, generally, they are different. By taking a glimpse at $\hat{\tau}_{IPW}$, the estimator weights the observed data by the inverse of the probability of receiving treatment or control. Intuitively, a subject represents larger population if he/she has lower chance of being sampled. In practice, it is common to normalize weights to reduce the variance of the weighting estimators, leading to a more stable estimate \citep{hirano2003efficient}. From the perspectives of semiparametric inference, $\hat{\tau}_{IPW}$ admits asymptotic linear expansion and reaches semi-parametric efficiency bound so that no regular estimator can improve its asymptotic performance, in other words, $\hat{\tau}_{IPW}$ is an optimal estimation of ATE \citep{hirano2003efficient, tsiatis2007semiparametric}. Even though IPW estimator $\hat{\tau}_{IPW}$ has good theoretical foundations, its unbiasedness and consistency highly rely on whether the propensity score model is correctly specified \citep{funk2011doubly}. Some researchers proposed using machine learning models to estimate propensity scores and \citet{dorie2019automated} has also showed the improvement by simulation compared with traditional logistic models, however, \citet{keele2021comparing} pointed out that black-box machine learning methods make little difference in real data. Additionally, similar to other propensity-score-based methods, $\hat{\tau}_{IPW}$ does not perform well when the estimated propensity scores are close to 0 or 1. To solve the issue of unstable IPW estimator causing by extreme weights, \citet{crump2009dealing} suggested further adjustment such as trimming the extreme weights to exclude subjects beyond the range of the common support region.

Doubly Robust (DB) Estimator, also called augmented IPW (AIPW), was firstly introduced by \citet{robins1994estimation} from the missing data perspective, which combines both regression-based approach and weighting method to provide double protection against misspecification \citep{lunceford2004stratification}: 
\begin{eqnarray*}
\hat{\tau}_{DR}&=& \hat{\tau}_{DR,1}-\hat{\tau}_{DR,0}\\
&=&\frac{1}{N} \sum_{i=1}^{N}\big[  \frac{Z_iY_i^{obs}}{\hat{e}(\bm{X}_i)} - \frac{Z_i - \hat{e}(\bm{X}_i) }{ \hat{e}(\bm{X}_i) } \cdot \hat{\mu}_1(\bm{X}_i)  \big] -  \frac{1}{N} \sum_{i=1}^{N}\big[  \frac{(1-Z_i)Y_i^{obs}}{1- \hat{e}(\bm{X}_i)} - \frac{Z_i - \hat{e}(\bm{X}_i)}{ 1-\hat{e}(\bm{X}_i) } \cdot \hat{\mu}_0(\bm{X}_i)  \big]\\
&=& \frac{1}{N} \sum_{i=1}^{N}\Big[  \hat{\mu}_1(\bm{X}_i) - \hat{\mu}_0(\bm{X}_i) \Big] + \frac{1}{N} \sum_{i=1}^{N}\Big[ Z_i \frac{Y_i^{obs}-\hat{\mu}_1(\bm{X}_i) }{\hat{e}(\bm{X}_i)} 
 - (1-Z_i) \frac{Y_i^{obs} - \hat{\mu}_0(\bm{X}_i) }{1- \hat{e}(\bm{X}_i)}  \Big].
\end{eqnarray*}
The DR estimator can be interpreted as estimating the PATE by using the regressions, then employing IPW to the residuals to adjust for bias. For this approach, we firstly estimates the nuisance parameters $\mu_z(\bm{x})$ and $e(\bm{x})$ non-parametrically, then estimates the parametric part,  average treatment effect. When either one of the postulated models is misspecifed, $\hat{\tau}_{DR}$ is still consistent for the PATE, which refers to as double-robustness property \citep{waernbaum2010propensity}. A sequence of papers have been developed to research on the properties of the DR estimator. \citet{glynn2010introduction} showed that the DR estimator is more stable when the propensity scores is close to 0 or 1 compared to IPW estimator. \citet{farrell2015robust} examined the behavior of DR estimator in high dimensional regression adjustments. \citet{robins1994estimation} laid the the foundation of the later research on the semiparametrically efficient property of DR estimator in estimating ATEs, and later, \citet{hahn1998role} demonstrated the effects of the propensity score for efficient semiparametric estimation on the ATEs and ATTs. In recent years, research on incorporating machine learning algorithms in estimating the nuisance part in DR estimator has become popular. \citet{westreich2010propensity} proposed using machine learning models such as neural networks, support vector machines, decision trees (CART), as alternatives to logistic regression in estimating propensity scores. Recently, several authors such as \citet{chernozhukov2018double, newey2018cross} have discussed estimating the nuisance parameters in DR estimator via cross-fitting so as to attain efficiency under some certain conditions. Specifically, the idea of cross-fitting is splitting the data into K-fold and use the holdout examples to estimate nuisance parameters, namely outcome regression and propensity scores, while the remaining examples are used in estimating the treatment effect. 

\subsection{Balancing Estimators in Inverse-Propensity Weighting}
Quite recently, considerable attention has been drew to the covariate balancing in estimating propensity scores. As pointed out by \citet{kang2007demystifying}, the major pitfall of IPW-based analysis is its sensitivity to misspecification of the propensity score model, which yields biased estimation on average treatment effect. In traditional propensity scores estimation, one can improve the fit by estimating propensity score via different models until the resulting covariate balance between treatment and controlled groups is satisfactory \citep{imbens2015causal}. \citet{imai2014covariate} introduced Covariate Balancing Propensity Score (CBPS) and advocated that since the essense of IPW analysis is to re-create a weighted population to make unconfounded comparison possible, one could use covariate balancing as a constraint to guide the propensity score estimation so that despite model misspecification, one can still achieve a good balance in covariate distributions between the treatment and control groups. Later, \citet{fan2016improving} discussed the optimal choice of balancing functions for the CBPS methodology. Several authors investigated the connections between covariate balance constraints and the dual convex optimization problem, including the work by \citet{hainmueller2012entropy, zubizarreta2015stable}. \citet{li2018balancing} proposed overlap weights that weights each subject by the probability of being assigned to the opposite treatment group, so as to achieve exact mean balance of all the included covariates and to solve the problems of extreme weights.  A landmark paper by \citet{zhao2019covariate} proposed a unifying framework that tailors these methods together via a loss function based approach. For high-dimensional inference on the ATE \citet{athey2016approximate} considered the use of augmented balancing estimators. Moreover, \citet{chernozhukov2016locally} generalized balance conditions into a broader use and developed the AIPW-based estimators for general functionals that can be written in terms of Riesz representer.


\subsection{Bayesian Paradigm in Causal Inference}
Potential outcomes framework can be view as a missing data problem intrinsically, since at most one of the potential outcomes can be realized for each subject. Bayesian modeling provides sophisticated frameworks in drawing inference from incomplete data, which have brought a lot of new insights into causal inference motivated by missing data perspectives.

Let $\big( \bm{Y}(1), \bm{Y}(0)\big)$ be the $N \times 2$ matrix  where $\bm{Y}(z)$ is the vector notation of potential outcomes under treatment $z$ with i-th entry equal to $Y_i(z)$, and let $\bm{Y}^{obs}$ and $\bm{Y}^{miss}$ be the vector notation with i-th entry equal to $Y_i^{obs}=Y_i(Z_i)$ and $Y_i^{obs}=Y_i(1-Z_i)$ respectively. In Frequentist paradigm, there is a subtle theoretical difference between the estimands of finite-sample ATE and super-population ATE. However, in Bayesian Paradigm, both observed $\bm{Y}^{obs}$ and unobserved $\bm{Y}^{miss}$ are treated as random variables, which results in a substantial difference between the estimation of these two estimands in the sense of the source of uncertainty. Specifically, \citet{ding2018causal} pointed out that for finite-sample estimands considering all potential outcomes as fixed values, Bayesian inference focus on imputing the missing potential outcomes based on the posterior draws, while super-population estimands considering all potential outcomes as random variables from a target distribution, $\bm{Y}^{obs}$ can also be simulated from the posterior predictive draws. 

Since a causal estimand that can be written as a function of $\tau=\tau\big( \bm{Y}(1), \bm{Y}(0),\bm{Z}, \bm{X}\big)$ can also be represented as $\tau=\tau\big( \bm{Y}^{obs},\bm{Y}^{miss}, \bm{Z}, \bm{X} \big)$,  research on methodologies in imputing $\bm{Y}^{miss}$ play an importation role in finite-sample inference \citep{imbens2015causal}. One of the most widely used Bayesian modeling was proposed by  \citet{rubin1978bayesian}. In his framework, beyond unconfoundedness assumption, the model also assumes exchangeability for the existence of the prior distribution $p(\bm{\theta})$, and distinct and independent prior parameters for $f(Z_i|\bm{X})$ and $f\big( Y_i(1), Y_i(0)|\bm{X}_i \big)$. The joint distribution of potential outcomes $f\big( Y_i(1), Y_i(0)|\bm{X}_i \big)$ is the key "model for science". $\bm{Y}^{miss}$ can be obtained from posterior predictive distribution through data augmentation \citep{tanner1987calculation}, also known as Gibbs sampling. For super-population causal estimand, both $\bm{Y}^{obs}$ and $\bm{Y}^{miss}$ are considered as random variables and can be simulated from the posterior predictive distributions. A conventional way of estimating population ATE in Bayesian inference is by using the empirical distribution of covariates $\bm{X}$ to compute the conditional ATE \citep{ding2018causal}.

The problem appears in both Frequentist or Bayesian framework, that only the marginal distributions of the potential outcomes can potentially be observed, but the correlation between potential outcomes,  denoted as $\rho$, is not observable. Some researchers might cast doubt on the assumption of independence between potential outcomes. In Frequentist framework, if the causal estimands are comparisons of potential outcomes under different treatment assignments, it is sufficient to estimate $\mathbb{E}\big[ Y_i(z) \big]$ without taking into account the non-identifiable parameter $\rho$. On the other hand, researchers might care the association between potential outcomes in some scenarios, for instance, when outcome of interests are ordinal, the estimand such as $P\big(Y_i(1)>Y_i(0)\big)$ might be of interest. From Bayesian's perspectives, even though all parameters are identifiable if imposing proper prior distributions \citep{lindley1972bayesian}, some parameters might be weakly identifiable if the posterior distributions are highly sensitive to the prior \citep{gustafson2015bayesian, ding2018causal}. To address this issue, \citet{richardson2011transparent} advocated transparent parametrization to isolate the non-identifiable parameter $\rho$ and treat it as a fixed value when simulating posterior draws. Later, \citet{ding2016potential} proposed a sensitivity analysis for the non-identifiable parameter $\rho$, and illustrated the idea using a completely randomized experiment with binary outcome of interest.

\subsection{Instrumental Variables Approach}
\begin{figure}[h]
    \centering
    \includegraphics[width=0.5\textwidth]{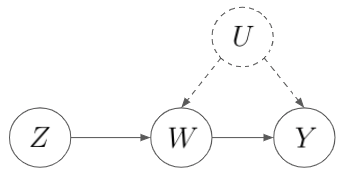}
    \caption{A Example of Instrumental Variable ($Z$: instrumental variable, $W$: treatment variable, $Y$: outcome of interest, and $U$: unmeasured confounders)}
    \label{fig:iv}
\end{figure}

Traditionally, unbiasedly estimating causal effects has primarily been based on the assumption of unconfoundedness. However, in some certain circumstances, there are variables, so-called instrumental variables (IV), that might potentially affect the treatment assignment so that indirectly affect the outcome of interests. For instance, the variable $Z$ in Figure \ref{fig:iv} can be considered as an instrumental variable that does not have a direct effect on outcome of interest $Y$ but affects the outcome through the variable $W$. As such, the unconfoundedness assumption seems implausible.

The IV approach has long been discussed in econometric literature, dominating by structural equation models that relies on linear parametric specifications to model the constant treatment effects \citep{wright1921correlation, haavelmo1943statistical, morgan1990history, stock2003retrospectives}. Since the influential work by  \citet{imbens1994identification}, \citet{angrist1996identification}, and \citet{imbens1997estimating}; that embedded instrumental variables into the potential outcomes framework and formulated the assumptions in a more traceable way, there has been a growing interest in IV method in recent statistics literature. A common IV estimand in statistic literature is Local Average Treatment Effect (LATE), introduced by \citet{imbens1994identification}, which is interchangeably referred to as Complier Average Treatment Effect (CATE) \citep{imbens1997estimating} since the inference is based on the average effect of the subpopulation of compliers. The finite-sample LATE is defined in Equation \ref{scate} below
\begin{equation*}\label{scate}
\tau_{SLATE}=\frac{1}{N_{co}} \sum_{i: G_i=co}\big[ Y_i(1)-Y_i(0) \big],
\end{equation*}
where $N_{co}=\sum_{i=1}^{N}\mathbbm{1}(G_i=co)$.

Although economists and statisticians approach the matter from different starting point, recent researches have shown that they shared a lot of similarities. For instance, \citet{imbens2014instrumental} gave a comprehensive review on the connections between the statistical literature on IV method and the econometrics point of views, and how they complemented each other along the way.

One popular application using instrumental variables is the estimation of treatment effect under noncompliance. Even though randomized experiments are the gold standard in establishing causal link, the presence of noncompliance with the assignment breaks the initial randomized assignment mechanism. A strand of the literature tackles the problem of noncompliance in design phase, such as \citet{zelan1979new, torgerson1998zelen}. Several naive methods have also been developed for randomized trials with noncompliance. For instance, as-treated analysis compares units based on actual treatment received; per-protocol analysis focuses only on the compliers and discarding complying units; and intention-to-treat (ITT) analysis compares units based on their initial random assignment \citep{sommer1991estimating, mcnamee2009intention}. 

A key limitation of the conventional ITT analysis is that it is centered mainly on the causal effects of the assignment on the outcome \citep{lee1991analysis}. However, the choices on the receipt of treatment are self-selected, which might capture the latent characteristics of each unit. In this sense, the causal effects of the treatment received might seem to be of more interest than the causal effects of the treatment assigned. IV approach provides an alternative that not only allows researchers to relax the assumption of unconfoundedness, but also enables the estimation of the effects on the receipt of treatment.

For the basic setup of randomized experiment with noncompliance, in addition to the binary treatment assignment $Z_i$ and outcome of interest $Y_i^{obs}$ as in the usual potential outcomes models, the actual receipt of treatment $W_i^{obs}=W_i(Z_i)$ is also observed. Each unit is associated with two potential outcomes on the treatment received, $W_i(0)$ and $W_i(1)$, so that they can be partitioned into subgroups as compliers (co), nevertakers (nt), defiers (df), and alwaystakers (at), based on their compliance status $G_i$, which is deterministic according to the pairs of values of $\big( W_i(0), W_i(1) \big)$  \citep{angrist1996identification}. More specifically, the four compliance types are defined as follows:
\begin{equation*}
C_i= \begin{cases}
\text{Nevertakers (nt)} & \text{if } W_i(0)=0, W_i(1)=0\\
\text{Compliers (co)} & \text{if } W_i(0)=0, W_i(1)=1\\
\text{Defiers (df)} & \text{if } W_i(0)=1, W_i(1)=0\\
\text{Alwaystakers (at)} & \text{if } W_i(0)=1, W_i(1)=1
\end{cases}. 
\end{equation*}
Because of the fundamental problem of causal inference that $W_i(Z_i)$ and $W_i(1-Z_i)$ are not jointly observable, the possible compliance types can not be inferred directly by the observed assignment $Z_i$ and the treatment received $W_i^{obs}$ without making extra assumptions. The key assumptions in identifying the estimand CATE are SUTVA, cross-world counterfactual independence, the exclusion restriction, and monotonicity. Under exclusion restriction assumption, $Y_i(z,w)$, denoting the double-indexed potential outcomes of the primary outcome of interest, can be written as $Y_i(w)$. CATE can be nonparametrically estimated by the method-of-moment-based estimator, the ratio of the ITT effects of primary outcome and treatment received \citep{angrist1996identification, imbens2015causal}. Later, \citet{frangakis2002principal} referred the compliance types as principal stratum and generalized noncompliance as a special case of post-treatment variable in their principal stratification approach. Apart from noncompliance, principal stratification method has been widely used in other applications such as censoring by death \citep{rubin2006causal,zhang2009extensions}, fuzzy regression discontinuity designs (FRD) \citep{hahn2001identification, chib2016bayesian}, and mediation analysis \citep{elliott2010bayesian}.

The main pitfall of moment-based estimator is its difficulty in including covariates. Therefore, \citet{imbens1997bayesian} and \citet{hirano2000assessing} outlined Bayesian inference for principal stratification that can incorporate  pre-treatment covariates in analysis. For each subject, only one of the potential outcomes, namely $W_i^{obs}=W_i(Z_i)$ and $Y_i^{obs}=Y_i(W_i^{obs})$, can be possibly observed, while $W_i^{miss}=W_i(1-Z_i)$ and $Y_i^{miss}=Y_i(1-W_i^{obs})$ are missing. Intrinsically, causal inference can be seen as a missing data problem. Let $\bm{Y}^{obs}, \bm{Y}^{miss}, \bm{W}^{obs}$, and  $\bm{W}^{miss}$ be the N-vector of observed and missing outcomes and receipt of treatment respectively. The goal in Bayesian paradigm is to derive the predictive distribution of $(\bm{Y}^{miss}, \bm{W}^{miss})$ based on the observed data $(\bm{Y}^{obs}, \bm{W}^{obs}, \bm{Z}, \bm{X})$ so as to compute the estimand $\tau_{SLATE}=\tau(\bm{Y}^{obs}, \bm{Y}^{miss}, \bm{W}^{obs}, \bm{W}^{miss}, \bm{Z}, \bm{X})$ as a function of observed and missing variables.

The core inputs that are needed to be specified in the model-based approach are the distribution of the compliance types $f(G_i|\bm{X}_i,\bm{\theta})$ and the joint distribution of potential outcomes
$f\big( Y_i(0), Y_i(1)\big| W_i(0), W_i(1),X_i; \bm{\theta} \big)$, or equivalently, $f\big( Y_i(0), Y_i(1)\big| G_i, X_i; \bm{\theta} \big)$. \citet{rubin2010bayesian} suggested using the two binary models, one for being a complier or not, and one for being a nevertaker conditional on not being a complier, to model the three-valued compliance status indicator. The main idea of this framework is that the observed pairs $(Z_i, W_i^{obs})$ consists of a mixture of subjects from different compliance types, so that methods like EM algorithm or Data Augmentation (DA), that are typically used in mixture model inference, can be used in inferencing causal effects in principal stratification \citep{imbens2015causal}. For DA method, \citet{richardson2011transparent} further pointed out that posterior is sensitive to the model specification of compliance types and proposed transparent parametrization by separating the identified and non-identified parameters.


\section{Causal Graphical Models} 

Potential outcome framework is powerful in recovering the effect of causes. In potential outcome framework, causal effects are answered by specific manipulation on treatments \citep{holland1986statistics}. However, when it comes to identifying the causal pathway or visualizing causal networks, potential outcome model have it own limitations. As an alternative, causal graphical models provides an intriguing tool in representing causal effects by directed edges that describes dependency relationships, which not only allows us to model interventions more generally, but also describes the data generating mechanism of the random variables.

\subsection{Path Analysis}
Functional causal models have a long history, tracing back to the original path analysis developed by geneticist \citet{wright1918nature}. They have been widely used in modeling the causal structures by combining the structural equation models (SEMs) and directed graphs \citep{wright1934method, pearl2009causality, pearl2018book}. Consider the random vector $\bm{X}_{[p]}=(X_1,...,X_p)$, the linear SEM consists of a set of equations of the form
\begin{equation}\label{linearSEM}
X_i = \beta_{0i}+\sum_{j \in pa(X_i)} \beta_{ji} X_j + \epsilon_i, \qquad i=1,2,...,p
\end{equation}
or equivalently,
\begin{equation}
\bm{X}_{[p]} = \bm{\beta}_0 + \bm{M}'\bm{X}_{[p]}+\bm{\epsilon}_{[p]}
\end{equation}
where $pa(X_i)$ denotes the set of variables that are parents (or direct predecessors) of $X_i$, $\epsilon_1,...,\epsilon_p$ are mutually independent noise terms with zero mean, $\beta_{ji}$'s are so-called path coefficients that quantify the causal effects of $X_j$ on $X_i$, and $\bm{M}$ is the matrix of path coefficients. This framework is structural since even if we intervene on several variables, the functional form among variables in equation (\ref{linearSEM}) would not change. The random variables $\bm{X}_{[p]}$ that satisfies the model structure of the form in Equation (\ref{linearSEM}) can be represented by a directed acyclic graph (DAG) $\mathcal{G}=(V,E)$, where V is the set of associated vertices, each corresponding to one of variable of interest $X_i$, and $E \subseteq V \times V$ is the corresponding edge set. Because for any DAG, the graph contains no cycles and there exists a topological ordering of the vertices, the matrix $I_d-\bm{M}$ is invertible and thus such structural assignments induces a unique solution $\bm{X}_{[p]}=(I_d-\bm{M})^{-1}(\bm{\beta}_0 + \bm{X}_{[p]})$.

Path analysis can be used in differentiating correlation and causation.  Applying Wright's path analysis, when the variance of $X_i$'s are all standardized, the covariance between $X_i$ and $X_j$ is the product of path coefficients that are on the edges of all d-connected paths, that is
\begin{equation}\label{linearSEMcov}
    Cov(X_i,X_j) = \sum_{(d_0,...,d_m) \in D(i,j)} \prod_{k=1}^{m}\beta_{d_{k-1}d_k},
\end{equation}
where $D(i,j)$ is the collection of the d-connected paths between $i$ and $j$. Additionally, in a linear SEM, causal effect of $X_i$ on $X_j$ can be defined as 
\begin{equation}
c(X_i \rightarrow X_j) = \sum_{(d_0,...,d_m) \in G(i,j)} \prod_{k=1}^{m}\beta_{d_{k-1}d_k},
\end{equation}
where $G(i,j)$ is the collection of the directed paths between $i$ and $j$. As can be seen from Equation (\ref{linearSEMcov}), if $Cov(X_i,X_j) \neq 0$ then there exists a d-connected path between $X_i$ and $X_j$. The d-connected path between $X_i$ and $X_j$ only introduces dependency between the two variables, which does not imply causation. The correlation between $X_i$ and $X_j$ implies causation only if $D(i,j)=G(i,j)$, in other words, all the d-connected paths between $i$ and $j$ are directed paths.

With the underlying graph structure given, confirmatory factor analysis focuses on the statistical inference part. Suppose the random vector $\bm{X}_{[p]}$ satisfies the linear SEM with Gaussian errors and the matrix of path coefficients $\bm{M}$ is identifiable, generally, $\bm{M}$ can be estimated by maximum likelihood estimator or generalized method of moments \citep{browne1984asymptotically}. Linear SEMs are also useful in identifying the causal effects between unobserved variables using "three indicator rule" \citep{o1994identification}. With pre-specified DAG and assumptions on the latent variables, the path coefficients between the latent variables are identifiable \citep{kuroki2014measurement}.

\subsection{Bayesian networks}
Causal inference can be naturally embedded in graphical models frameworks since the dependencies and interactions between variables can be presented by graphs with probabilistic distributions, in which nodes correspond to variables of interests and edges represents associations. In Bayesian networks, causalities among variables are holistically represented in the form of graphs with directed paths carrying causal information. A Bayesian network is a collection of directed graphical models, which can be described using directed acyclic graphs (DAG), denoted by $\mathcal{G}=(V,E)$, where $V$ is a finite set of vertices $i$'s corresponding to a set of random variables $\bm{X}_{[p]}=\{X_1,...,X_p\}$, i.e. $i \mapsto X_i$, and $E \subseteq V \times V$ is the corresponding edge set. A DAG is a directed graph with no cycles. A node $j$ in a DAG graph $\mathcal{G}$ is a parent of node $i$ if there exists directed edge $(i,j)$ such that $j \rightarrow i$. In addition to connecting two variables using directed edges, two variables in a DAG can be connected by different paths in the graph, where path is a sequence of adjacent nodes with directed arrows that connect those two variables.

A DAG encodes dependency relationships by associating each node with a conditional probability distribution of the corresponding variable $x_i$ given its parents' values $\bm{x}_{pa(i)}$, where $pa(i)$ denotes the set of parental nodes of the vertex $i$. A joint probability distribution $\mathbb{P}$ factorizes with respect to a DAG $\mathcal{G}$ if it satisfies:
\begin{equation}\label{conditional_prop_dist}
    f(x_1,...,x_p) = \prod_{i} f(x_i|\bm{x}_{pa(i)}).
\end{equation}

Bayesian networks provide a natural setup for expressing dependency relationships. It models the changes in joint distribution related to external interventions. As an example, consider a model of four variables, in which variable $X_1$ affects both variables $X_2$ and $X_3$, and variable $X_4$ depends on the values of $X_2$ and $X_3$. Based on the dependency assumptions, this Bayesian network defines a joint probability distribution as follows:
\begin{equation*}
f(x_1,x_2,x_3,x_4)=f(x_1)f(x_2|x_1)f(x_3|x_1)f(x_4|x_2,x_3).   
\end{equation*}
\begin{figure}[h]
    \centering
    \includegraphics[width=0.3\textwidth]{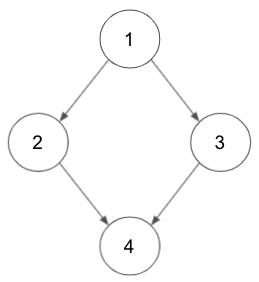}
    \caption{An Example of a DAG}
    \label{fig:DAG_example}
\end{figure}
Figure \ref{fig:DAG_example} visualizes the dependencies among those variables in a DAG.

With structure given by a DAG $\mathcal{G}$, the associated joint probability distribution can be expressed compactly by conditional distributions. For instance, without knowing the relationships among variables, a naive approach in describing a probability distribution with $p$ binary random variables requires $O(2^p)$ parameters. Given DAG structure that the maximum number of incoming edges is $m$, the size of parameters in describing the entire probability distribution can be reduced to $O(p\cdot 2^m)$ \citep{russell2002artificial}. The compact expression also induces certain independence assumptions in the model. 

Several criteria have been developed for reading-off conditional independencies from DAGs. The global Markov property \citep{pearl1986constraint, lauritzen1990independence}, together with the Hammersley-Clifford theorem \citep{cowell1999building}, provide a tool in reading-off conditional independencies for undirected graph. For DAGs, the Hammersley-Clifford theorem can not be directly applied to recover independencies among variables, since the densities in general do not factorize with respect to cliques (a subset of nodes that every two nodes are directly connected in the undirected graph). The undirected moral graph $\mathcal{G}^m$ of a DAG $\mathcal{G}$, obtained by adding edges between parents having common child, suggests an alternative in checking factorization with respect to a DAG \citep{cowell1999building}. The theorem states that if a probability distribution factorizes with respect to a DAG $\mathcal{G}$, it also factorizes with respect to the $\mathcal{G}^m$. Then the Hammersley-Clifford theorem can be used to check whether the $\mathcal{G}$ satisfies the global Markov property so as to obtain conditional independencies in the original DAG.

\begin{figure}[h]
    \centering
    \includegraphics[width=0.25\textwidth]{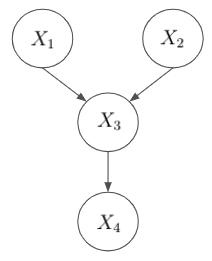}
    \caption{An Example of a Collider in a DAG}
    \label{fig:swig_7}
\end{figure}

The d-separation rule is also a criterion in obtaining conditional independencies for DAGs. The criterion emphasizes the role of colliders in introducing dependencies. A node is said to be a collider if it is a common child of the other two or more nodes. For instance, in Figure \ref{fig:swig_7}, the variable $X_3$ is a collider, since it is directed influenced by the variables $X_1$ and $X_2$. Conditioning in general removes dependencies between variables. As can be seen from this example, the variables $X_1$ and $X_2$ are originally independent, i.e. $X_1 \perp X_2$. However, when conditioning on the collider variable $X_3$ or the descendent of the collider variable, conditioning introduces dependencies between variables, i.e. $X_1 \not\!\perp\!\!\!\perp X_2 \big|X_3$ and $X_1 \not\!\perp\!\!\!\perp X_2 \big|X_4$.

In this sense, the d-separation rule states as the followings. In a DAG $\mathcal{G}=(V,E)$, a path is blocked by the set $C \subset V$ if the path contains a non-collider node $c_1 \in C$, or it contains a collider node $c_2 \notin C$ and $des(c_2) \cap C = \emptyset$, where $des(c_2)$ denotes the set of the descendants of the node $c_2$ \citep{pearl1991theory}. For any sets $V_1,V_2,V_3 \subset V$, if every path from a node in $V_1$ to a node in $V_2$ is blocked by $V_3$, then $V_1$ and $V_2$ are d-separated by $V_3$, denoted as $V_1 \perp V_2 | V_3 \; [\mathcal{G}]$. The theorem says that in a DAG $\mathcal{G}=(V,E)$ with respect to a probability distribution of $\bm{X}_{[p]}$, for any disjoint sets $V_1,V_2,V_3 \subset V$, the d-separation of $V_1$ and $V_2$ by $V_3$ implies conditional independencies i.e. $\bm{X}_{V_1} \perp \bm{X}_{V_2} | \bm{X}_{V_3}$, if and only if the probability distribution factorizes with respect to the DAG. In other words, a probability distribution with respect to the DAG satisfies the global Markov property if the d-separation criterion holds. Compared with constructing intermediate moral graphs, the d-separation criterion does not require the probability distribution to have positive density, and can be directly applied to original DAGs to read-off independencies.

\subsection{Non-parametric Structural Equation Models (NPSEMs)}
Linear SEMs impose a strong assumption on linearity. \citet{pearl1991theory} adapted the structural equation models and relaxed linearity assumptions by introduction non-parametric structural equation models (NPSEMs). For NPSEMs, the arrows in the DAGs represent functional relationships instead of inducing linearity. Formally, random variables $\bm{X}_{[p]}=(X_1,...,X_p)$ with respect to a DAG $\mathcal{G}=(V,E)$ can be further expressed as a deterministic function
\begin{equation}\label{npsem}
    X_i=f_i\big( \bm{X}_{pa(i)}, \epsilon_i \big), \qquad i=1,...,p.
\end{equation}
When further assuming the noise terms, $\epsilon_i$'s, to be mutually independent, the model is referred to as non-parametric structural equation models with independent errors (NPSEM-IE). In NPSEMs framework, the sampling distributions of random variables can be expressed by both equations (\ref{conditional_prop_dist}) and (\ref{npsem}), while the latter describes the whole data generating mechanism and provides an essential tool for reasoning causal effects through the concept of do-calculus \citep{pearl1995causal, pearl1998graphical, pearl2000causality}.

Their functional mechanism $X_i=f_i\big(pa(X_i), \epsilon_i \big)$ provides a general method in querying causal effects when there are exogenous interventions on several variables. One of the most famous tool in answering such causal queries is the "do calculus" advocated by \citet{pearl1995causal}. The concept of do-calculus is as follows. The functional characterization in equation (\ref{npsem}) defines how the values of $X_i$'s would change if external interventions were made to some of the variables in the system. Specifically, for any two subsets of nodes $\bm{X}_A, \bm{X}_B \subseteq \bm{X}_{[p]}$, where $\bm{X}_A \cap \bm{X}_B = \emptyset$, when making an intervention of setting $\bm{X}_A$ to certain values $\bm{x}_A$, the structural equations can be modified accordingly by replacing the equations with $\bm{X}_A=\bm{x}_A$ and the causal effect on $\bm{X}_B$ can be estimated using the newly computed observables by $\mathbb{E}\big[ \bm{X}_B \big| do(\bm{X}_A=\bm{x}_A) \big]$.

The counterfactuals from a DAG can be also modeled by using NPSEMs. For intervention of setting $\bm{X}_A=\bm{x}_A$, where $A \subseteq [p]$, the counterfactual variables $\{X_i(\bm{X}_A=\bm{x}_A) \, \big| \, i \in [p] \}$, or abbreviated as $\{X_i(\bm{x}_A) \, \big| \, i \in [p] \}$, can be defined by recursive substitution \citep{malinsky2019potential}. Specifically, for the interventions on the parental nodes of $X_i$, i.e., setting $\bm{X}_{pa(i)}=\bm{x}_{pa(i)}$, the counterfactual of $X_i$ is defined as 
\begin{equation}
    X_i(\bm{X}_{pa(i)}=\bm{x}_{pa(i)}).
\end{equation}
On the other hand, if the interventions are made on the ancestor set $A$, i.e. $A \neq pa(i)$, the counterfactual is defined as
\begin{equation}
    X_i(\bm{X}_A=\bm{x}_A)=X_i\Big( \bm{X}_{pa(i) \cap A} = \bm{x}_{pa(i) \cap A},  \bm{X}_{pa(i) \setminus A} = \bm{X}_{pa(i) \setminus A} (\bm{x}_A) \Big).
\end{equation}
Defining in this way, it implies the consistency property \citep{malinsky2019potential}: for any disjoint sets $A,B \subseteq V$, $i \in V \setminus (A \cup B)$,
\begin{equation}
    \bm{X}_B(\bm{x}_A)=\bm{x}_B \text{ implies } X_i(\bm{x}_A, \bm{x}_B)=X_i(\bm{x}_A).
\end{equation}
The intuition behind the consistency property is that if the system is intervened by setting the values to what they would have in the first place, then this additional intervention would not make a difference to the system.

\begin{figure}[h]
    \centering
    \includegraphics[width=0.35\textwidth]{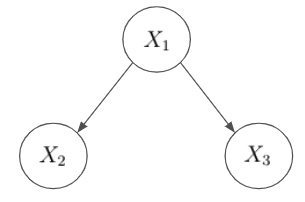}
    \caption{An Illustration for Multiple-world and Single-world Independence Assumptions}
    \label{fig:independence_assumption}
\end{figure}

Multiple-world independence assumption assumes the variables $\{ X_i(\bm{x}_{pa(i)})| \bm{x}_{pa(i)}\}$ are mutually independent for all $i \in [p]$, while single-world independence assumption assumes the variables $\{ X_i(\bm{x}_{pa(i)})\}$ are mutually independent for all $i \in [p]$ \citep{richardson2013single}. Intuitively speaking, multiple-world independence makes hypothesis on the potential outcomes that can never occur simultaneously, while single world independence assumes that all of the potential outcomes are independent of each other if the specified variables were intervened in the graph. For instance, consider the graph structure in Figure \ref{fig:independence_assumption} with binary variable $X_1$. The multiple-world independence makes assumptions on the counterfactual of $X_1$ that can never occur simultaneously, i.e. $X_2(X_1=1) \perp X_3(X_1=0)$, while the single-world independence only requires $X_2(X_1=0) \perp X_3(X_1=0)$.
NPSEM-IE framework induces multiple-world independence assumption in the model by assuming independence of $\epsilon_i$'s. In general, the multiple-world independence assumption is non-testable since the assumption is based on the counterfactuals that can never happen at the same time.

\subsection{Single-world Intervention Graphs (SWIGs)}
\begin{figure}[h]
    \centering
    \includegraphics[width=0.55\textwidth]{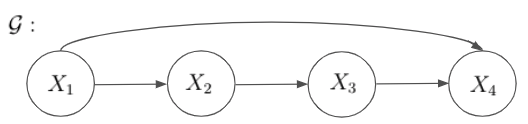}
    \caption{A Graphical Structure for Illustration of SWIG}
    \label{fig:swig1}
\end{figure}

\citet{richardson2013single} argue that multiple-world independence is a strong assumption for NPSEM, and to model counterfactual variables in a graphical model, single-world independence assumption is suffices for causal identification. They introduced the single-world intervention graph (SWIG) that unifies the graphical theories and potential outcomes framework. Specifically, for any intervention of setting $\bm{X}_A=\bm{x}_A$ in the system, the SWIG, denoted as $\mathcal{G}[\bm{X}(\bm{X}_A=\bm{x}_A)]$, can be constructed from a causal DAG $\mathcal{G}$ by splitting all of the vertices in $A$ into a random and a fixed component, then re-labelling each random node $X_i$ as $X_i(\bm{x}_{A \cap an(i)})$. Next, we are going to illustrate the concept of how a SWIG can be constructed for the graphical model in Figure \ref{fig:swig1}.

\begin{figure}[h]
    \centering
    \includegraphics[width=0.6\textwidth]{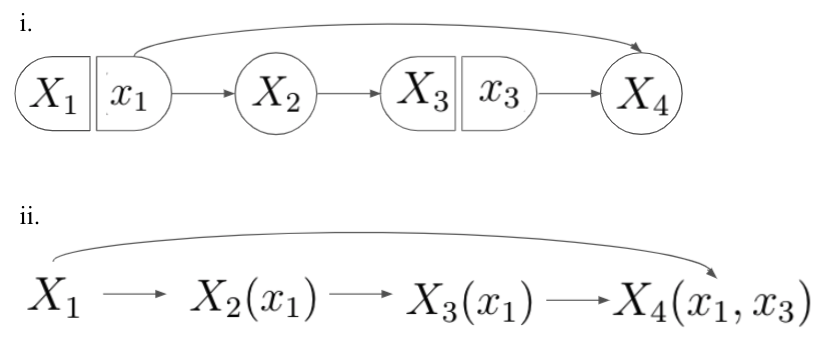}
    \caption{The SWIG of the Graph Structure in Figure \ref{fig:swig1} for intervention $(X_1=x_1,X_3=x_3)$}
    \label{fig:swig2}
\end{figure}

Consider a specific intervention $(X_1=x_1,X_3=x_3)$ made in the system. In the first step of constructing SWIG, all the nodes in the intervention set $\{X_1,X_3\}$ are split into a random and a fixed part, with random components getting all the incoming edges while fixed components getting all the outgoing edges. Then in the second step, all the random vertices in the graph from first step are relabeled. The Figure \ref{fig:swig2} shows the SWIG for the graphical model in Figure \ref{fig:swig1}, denoted as $\mathcal{G}\Big( \bm{X}(X_1=x_1,X_3=x_3) \Big)$ or $\mathcal{G}\Big( \bm{X}(x_1,x_3) \Big)$. As can be seen from the SWIG, the variables $X_2(x_1)$, $X_3(x_1)$, and $X_4(x_1,x_3)$ now becomes counterfactual variables.

In SWIGs, it consists of both random and fixed components. By conditioning on the fixed parts, one can construct subgraph $\mathcal{G}^{*}\big[\bm{X}(\bm{x}_A)\big]$ inherited only the random parts from the SWIG  $\mathcal{G}\big[\bm{X}(\bm{x}_A)\big]$, then apply d-seperation criterion \citep{pearl1988probabilistic} to read-off conditional independence among counterfactual variables. Alternatively, in the case of no unmeasured variables, g-computation formula \citep{robins1986new, pearl2009causality} can also be used to derive identification formulas for counterfactuals. Specifically, for continuous variable $\bm{X}_{[p]}$ satisfying single-world causal model with respect to a DAG $\mathcal{G}=(V, E)$, for any disjoint sets $A,B \subseteq V$,
\begin{equation}\label{g_formula}
    \mathbb{P}\big( \bm{X}_{B}(\bm{x}_{A}) = \Tilde{\bm{x}}_{B} \big) = \int \prod_{i \in  V \setminus A} \mathbb{P}\big( X_i = \Tilde{x}_i \big| \bm{X}_{pa(i) \cap A} = \bm{x}_{pa(i) \cap A},  \bm{X}_{pa(i) \setminus A} = \Tilde{\bm{x}}_{pa(i) \setminus A}  \big)   d{\Tilde{\bm{x}}_{C}},
\end{equation}
where $C= V \setminus (A \cup B)$. The identification formula for discrete case can be derived analogously by replacing the integral with summation. If graphical structures contains hidden variables, the front-door criterion and back-door adjustment advocated by \citet{pearl1995causal}, and potential outcomes calculus proposed by \citep{malinsky2019potential} are commonly used in deriving identification formulas.

\subsection{Structure Discovery}
Structure learning, on the other hand, focuses on re-constructing graphs and discovering causal structures based on conditional independencies in observational data. The concept of faithfulness assumption was introduced by \citet{pearl1988probabilistic} and \citet{spirtes1993causation}, providing the theoretical foundation for a series of algorithms using independence tests to learn causal structures. A probability distribution with respect to DAG $\mathcal{G}$ is said to be faithful if $A \perp C| B \; [\mathcal{G}] \Leftrightarrow \bm{X}_{A} \perp \bm{X}_{C} | \bm{X}_{B}$ holds. The existence of faithfulness of probability distributions have been proved in the class of multivariate normal distributions and the class of multinomial distributions. \citet{meek1995strong} further proved that the faithful distributions exist for almost all the discrete distributions that are Markov to a DAG $\mathcal{G}$, except those with Lebesgue measure zero. Most of the conditional-independence-test-based algorithms that can identify graphs up to Markov equivalence class usually require the faithfulness assumption, which can be regarded as a relatively weak assumption in structure learning \citep{pearl2000causality}. Some state-of-the-art algorithms include PC algorithm, IC algorithm, and SGS algorithm \citep{pearl2000causality}. Those algorithms depend on statistical tests of conditional independence to remove a number of possible structures and output a set of DAGs with d-seperated variables. \citet{shah2020hardness} pointed out that conditional independence test can be a hard statistical hypothesis and proposed using a test statistic computed based on generalised covariance measure. The greedy equivalence search (GES) algorithm introduced by \citet{chickering2002optimal} takes another approach using score-based criterion. For Gaussian data, the algorithm grows a graph greedily from an empty graph by maximizing the score function, which in general provides more stable estimates of the Markov equivalence class of the DAGs than PC does.

\section{Conclusion}
The theoretical development of causal inference was driven by its diverse applications in observational data to understand the impact of treatments on outcome variables. The potential outcomes or counterfactuals framework is convenient for statistical inference and making causal argument. For observational analysis, this framework often requires non-verifiable assumptions such as unconfoundedness. Causal graphical models represent causal effects among variables using directed edges, which is powerful in visualizing the whole causal networks and learning causal pathways. With graphical structures given, a series of theories have been developed in reading-off conditional dependencies among variables from graphs. Structure discovery on the other hand provides a way to recover Markov equivalence class based on the independencies in observational data. It is useful for prioritization experiments or when intervention on all the variables are impractical. In recent years, research on unifying those methods has become popular. More research in bridging the perspectives of graphical models and counterfactual framework would be valuable.

\bibliographystyle{apalike}
\bibliography{thesis}
\end{document}